\documentclass{svproc}
\usepackage{url}
\usepackage{graphicx}        
\usepackage{amsfonts}
\usepackage{multirow}
\usepackage[linesnumbered,ruled,vlined]{algorithm2e}
\usepackage{cite}
\usepackage{amsmath}
\usepackage{amssymb}
\usepackage{comment}
\usepackage{xcolor}

\usepackage{unicode-math}

\usepackage{lscape}

\usepackage{multirow}

\usepackage{array}
\newcolumntype{P}[1]{>{\centering\arraybackslash}p{#1}}

\SetKwInput{KwInput}{Input}                
\SetKwInput{KwOutput}{Output}              

\begin{document}
\mainmatter              
\title{Modularity-based Backbone Extraction in Weighted Complex Networks}
\titlerunning{Modularity-based Backbone Extraction}  
%
\author{Stephany Rajeh* \and Marinette Savonnet  \and
	Eric Leclercq  \and
	Hocine Cherifi }
\authorrunning{S. Rajeh et al.} 
%
%

\institute{Laboratoire d’Informatique de Bourgogne - University of Burgundy, Dijon, France\\
	\email{*stephany.rajeh@u-bourgogne.fr}}

\maketitle    

\begin{abstract}
The constantly growing size of real-world networks is a great challenge. Therefore, building a compact version of networks allowing their analyses is a must. Backbone extraction techniques are among the leading solutions to reduce network size while preserving its features. Coarse-graining merges similar nodes to reduce the network size, while filter-based methods remove nodes or edges according to a specific statistical property. Since community structure is ubiquitous in real-world networks, preserving it in the backbone extraction process is of prime interest. To this end, we propose a filter-based method. The so-called ``modularity vitality backbone" removes nodes with the lower contribution to the network's modularity. Experimental results show that the proposed strategy outperforms the ``overlapping nodes ego backbone" and the ``overlapping nodes and hub backbone." These two backbone extraction processes recently introduced have proved their efficacy to preserve better the information of the original network than the popular disparity filter.
\end{abstract}

\section{Introduction}
Complex networks, such as communication, biological, transportation, and contact networks, are widely analyzed. The daily production of data results in tremendously large real-world networks. Consequently, the analysis of such networks containing millions of nodes and billions of edges is more and more challenging, if not impossible, due to memory and time constraints. Thus, suitable extraction of the pertinent nodes and edges that preserve the essential information while reducing the size of the network is fundamental. Network backbones offer a way to do so. Two main research paths tackle this problem: coarse-grain backbones or filter-based backbones. In the former, one clusters together nodes sharing similarities to reduce the network size\cite{zeng2011coarse,zeng2019new}. In the latter, one removes nodes or edges from the network based on a given property \cite{coscia2017backboning, 10.1093/comnet/cnab021}.
The community structure is one of the significant properties in real-world networks. Indeed, it heavily determines their dynamics and their underlying functionalities \cite{boccaletti2006complex}. It is generally illustrated by dense regions of connected nodes that barely connect from one region to another. Communities can be non-overlapping or overlapping \cite{jebabli2018community}, hierarchical \cite{peel2020detectability}, and attributed \cite{atzmueller2021mining}. Community detection is one of the most prolific research areas in network science. It relies on numerous measures quantifying the quality of the community structure. Modularity is among the most popular\cite{newman2006modularity}. It compares the density of connections of the uncovered community structure with a similar random network. The higher the modularity, the higher the confidence in the tight community structure of the network.

Recent works have shown that one can exploit the community structure efficiently to extract backbones \cite{ghalmane2020extracting, ghalmane2021extracting}. Inspired by these works, we propose a filtering technique based on the preservation of the community structure of the network. It exploits the community structure using the concept of vitality. Vitality quantifies the contribution of a node to a given quality measure by removing this node and computing the variation of the quality measure. To assess its importance, we compute modularity as a quality measure, with and without the node in question. Then, one ranks the nodes from the lowest contribution on modularity to the highest. Subsequently, nodes with the lowest contribution are removed until one reaches the desired size of the network. 

Comparative experimental evaluations are conducted on real-world weighted networks of different sizes and domains. The developed backbone extraction technique called ``modularity vitality backbone" is compared with the recently introduced community-based method ``overlapping nodes ego backbone" \cite{ghalmane2020extracting}. Results show that it is more effective in preserving the core information of the network and the community structure.
\newline
The main contributions of the paper summarize as follows:
\begin{itemize}
	\item We propose a backbone filtering technique exploiting the community structure of networks.
	\item Experiments with weighted networks show that it outperforms another alternative measure.
	\item It can be easily adapted to any type of network (i.e., undirected, unweighted, and directed networks).
\end{itemize}

The remaining of the article is organized as follows. Section \ref{sec:relatedworks} discusses briefly the related works.  Section \ref{sec:VCB} introduces the modularity vitality backbone. Sections \ref{sec:data} and \ref{sec:evaluationmeasures} present respectively the datasets and the evaluation measures used in this study. Section \ref{sec:experimentalresults} reports the results of the comparative evaluation. Section \ref{sec:discussion} discusses the results. Finally, section \ref{sec:conc} concludes the paper.

\section{Related Works}
\label{sec:relatedworks}
Backbones offer an ideal solution to the trade-off between preserving essential information in the network and reducing the network size. Backbone extraction studies concerns mainly two types of networks: mono-mode networks \cite{van2012high, cao2019motif, 10.1093/comnet/cnab021} and bipartite networks \cite{zweig2011systematic, neal2013identifying, neal2014backbone}. In this work, we consider mono-mode networks. Within this class of networks, there are two leading approaches for extracting backbones. The first is coarse-graining, and the second is filtering.

In coarse-graining methods, one group nodes with similar characteristics into a single node. For instance, authors in \cite{gfeller2007spectral} merge the nodes based on random walks. In the work of \cite{chen2016effectively}, authors use the $k$-nearest neighbors algorithm ($k$-NN) to group similar nodes based on the nearest higher-density-neighbor.

In filter-based methods, the goal is to remove redundant information by pruning nodes or edges in the network. Redundancy is assessed based on a statistical property. 

Most of the works reported in the literature concern edge-filtering techniques. Serrano et al. propose the disparity filter. It uses a null model of the edge weights to preserve statistically significant edges \cite{serrano2009extracting}. Authors in \cite{goh2006skeleton} compute the betweenness centrality of edges and remove the ones that don't exceed a specific threshold. Authors in \cite{zhang2018extracting} use a combination of local and global information to extract the backbone. More precisely, they use the link weights to build the $h$-strength graph and the betweenness centrality to build the $h$-bridge graph. Then one obtains the backbone by merging $h$-strength and $h$-bridge. Simas et al.  \cite{10.1093/comnet/cnab021} present the distance backbone based on the triangular organization of edges which preserves all shortest paths.

Node filtering techniques are less frequent. They rely on topological features to associate a score to the nodes. Nodes with the higher scores are then extracted \cite{DAI2018271, RePEc}.
In this line, in their recent work, Ghalmane et al. prune nodes based on the community structure characteristics of the network \cite{ghalmane2020extracting}. They propose two node-filtering techniques. The first one preserves the overlapping nodes and the hubs of the network. In contrast, the second conserves the overlapping nodes and their one-step neighbors to form the backbone. These two algorithms exhibit superior performances as compared to the popular disparity filter. These results illustrate the community structure's importance in preserving the core information in a network while reducing its size.
Inspired by these findings, we propose the ``modularity vitality backbone" algorithm. This node filtering technique also exploits the community structure of the network.  It uses a measure of the node contribution to the modularity. Roughly speaking, nodes with the lowest contribution to the quality measure of the community structure are filtered. The remaining nodes form the backbone.


\section{Modularity Vitality Backbone}
\label{sec:VCB}
This section presents the vitality concept. Then, we briefly discuss various mesoscopic quality measures. We explain why we choose Newman's modularity as a quality measure. Finally, the algorithm of the proposed backbone extractor is given.

\subsubsection{Vitality Index.}
Let $G(V,E)$ be a simple and undirected graph where $V=\{v_1, v_2, ..., v_N\}$ is the set of nodes totaling $N=|V|$ and $E=\{(v_i, v_j) | v_i, v_j \in V\}$ is the set of edges. Denote $f(G)$ and $f(G\setminus \{u\})$ as two real-valued functions defined on the complete graph $G$ and on graph $G\setminus\{u\}$ without node or edge $u$. Then, the vitality index is the difference between both functions, defined as $\nu(G, u) = f(G) - f(G\setminus \{u\})$. The resulting value is a signed value, indicating the positive or negative contribution of the node or edge $u$ on graph $G$ \cite{koschutzki2005centrality}.

\subsubsection{Mesoscopic Quality Measures.}
There are numerous quality measures to characterize communities \cite{yang2015defining, leskovec2010empirical, rajeh2021characterizing}. Their goal is to answer how good is the community structure in a network. They use topological properties defined at the mesoscopic level. Let set $C=\{c_1, c_2,...,c_l,..., c_{n_c}\}$ represent $n_c = |C|$ communities of a graph $G$ and $f(c_l)$ represents a quality function of community $c_l$. One can categorize the quality functions into three main groups: \newline
\textbf{1. Based on internal connectivity:} such as internal density characterizing how densely connected the nodes are in a community compared to other communities. \newline
\textbf{2. Based on internal and external connectivity:} such as Flake-ODF measuring the fraction of nodes in a community with fewer internal edges than external ones. \newline
\textbf{3. Based on a network model:} such as Newman's modularity \cite{newman2006modularity} which assesses the difference between the real connections in the community $c_l$ and the random connections in the same community.

All of these quality functions characterize a single community. Hence, to quantify the quality of the overall community structure, one averages $f(c_l)$ over all the communities.

Newman's modularity is one of the most popular mesoscopic quality measures. Indeed, it is widely used in community detection algorithms as an optimization criterion \cite{clauset2004finding, brandes2007modularity, blondel2008fast}. This is the main reason why it is one of the well-accepted benchmarks for characterizing the community structure of the networks. Numerous extensions have also been proposed for modularity to account for networks with overlapping and hierarchical community structure \cite{chen2015fuzzy}. In this work, we use Newman's modularity as a quality measure to assess the vitality of nodes due to the following reasons: \newline
1. Modularity can naturally be extended to unweighted, undirected, and directed networks. \newline
2. Modularity vitality ensures that nodes that are the main contributors to the community structure are retained, regardless of their type (i.e., hub- or bridge-like).\newline
3. Previous works on modularity vitality centrality has proved to assign high scores to the most influential nodes \cite{magelinski2021measuring}.

\subsubsection{Algorithm.}
The ``modularity vitality backbone" is based on the vitality concept, where one can measure the contribution of a node or an edge using any quality measure computed on graphs. We use Newman's modularity as a quality measure. Nonetheless, one can opt for other quality measures to quantify the node and edge influence.

Newman's modularity enables us to differentiate between highly internally connected nodes (hubs) and nodes at the borders of the communities (bridges). Indeed, hubs increase the internal density of the communities. Therefore, they contribute positively to modularity. In contrast, bridges increase the connections between the communities. Consequently, they tend to decrease the modularity. As we choose to give equal importance to both types of nodes, we rank the nodes according to the absolute value of their modularity vitality score. It allows keeping nodes with the highest contribution to modularity, regardless of their role (i.e., hub nodes or bridge nodes). Then, one removes the nodes that barely contribute to modularity. The backbone extraction procedure is given in Algorithm 1.

\begin{algorithm}[!ht]
\DontPrintSemicolon
  
  \KwInput{Graph $G(V,E)$, Community set: $C=\{c_1,c_2, ..., c_{n_c}\}$, Size $s$}
  \KwOutput{Pruned graph $\hat{G}(\hat{V},\hat{E})$}

  $Q(G) \leftarrow Modularity (G, C)$ \tcp*{Computing modularity vitality of nodes}
  $D \leftarrow \varnothing$ \;
  \For{$v \in V$}{
  $Q(G \setminus \{v\}) \leftarrow   \sum_{c \in C}  \left[ 
    \frac{|E_{c}^{in}|-|E_{v}^{in}|}{|E|-|E_{v}|} - \left(
    \frac{2\left(|E_{c}^{in}| - |E_{v, c}^{in}|\right) + \left(|E_{c}^{out}| - |E_{v, c}^{out}| \right)}{2(|E|-|E_{v}|)} \right)^2\right]$\;
            
  $\alpha(v) \leftarrow Q(G) - Q(G \setminus \{v\})$\;
  $D[v] = |\alpha(v)|$
    }
  
  $D \leftarrow sort(D)$  \;
   \While{$|V|> s$}
   {  
   $\eta \leftarrow D.pop(v)$ \tcp*{Extracting the backbone}
   	$G \leftarrow G \setminus \eta$ \;
    $V \leftarrow V \setminus \eta$

  \If{$G$ is disconnected}
    {
        $G \leftarrow LCC(G)$
    }
    
   }

\caption{Modularity Vitality Backbone Extraction}
\end{algorithm}

Note that the vitality computation is not naively computed two times for each node. In such a case, the complexity can rapidly become prohibitive. Indeed, one computes instead the modularity variation reducing the computation's complexity to $O(|E|+Nn_c)$. It makes the vitality measure suitable for large-scale weighted networks. We also note that the symbol $|E|$ is extended to weighted networks.

\section{Datasets}
\label{sec:data}

We use a set of seven real-world networks originating from various domains (social, collaborative, ecological, and technological) in the experiments. The nodes and edges range from hundreds to thousands. We choose to integrate small networks in the experiments in order to get a better understanding on the filtering process. Table \ref{TableCharacteristcs} presents their basic topological characteristics. All the networks are freely available online\footnote{Aaron Clauset, Ellen Tucker, and Matthias Sainz, ``The Colorado Index of Complex Networks." https://icon.colorado.edu/ (2016).}\footnote{Tiago P. Peixoto, ``The Netzschleuder network catalogue and repository," https://networks.skewed.de/ (2020).}. As there is no ground truth available, we rely on the Louvain community detection algorithm to uncover their community structure \cite{blondel2008fast}.
\newline
\textbf{1. Zachary's Karate Club:} Nodes are members of a karate club and are connected if they are friends inside and outside the club. Edges are weighted by the relative interactions occurring between the members.\newline
\textbf{2. Wind Surfers:} Nodes are windsurfers in southern California in the fall of 1986. They are connected if they're friends. Edges are weighted based on the social closeness of the surfers to one another. \newline
\textbf{3. Madrid Train Bombing:} Nodes are terrorists in the train bombing of March 11, 2004, in Madrid. Edges represent contacts between the terrorists and are weighted based on the strength of their underlying relationship.\newline
\textbf{4. Les Mis\'erables:} Nodes are the characters in the novel ``Les Mis\'erables." Edges represent characters' co-appearances in the same chapter. They are weighted by the number of co-appearances.\newline
\textbf{5. Wiki Science:} Nodes are either applied, formal, natural, or social sciences Wikipedia pages. They're weighted by the cosine similarity between them. \newline
\textbf{6. Unicode Languages:} A bipartite network representing languages and countries. Weights represent the fraction of people in a given country having the literacy (reading and writing) of a specific language. \newline
\textbf{7. Scientific Collaboration:} Nodes are authors of articles in the ``Condensed Matter" category of arXiv. Edges represent co-authorship and are weighted by the number of joint papers among the authors.

\begin{table}[h]
   \centering
    \caption{Basic topological properties of the real-world networks under study. \textit{N} is the number of nodes. $|E|$ is the number of edges. $<k>$ is the average weighted degree. $\omega$ is the density. $\zeta$ is the transitivity. $k_{nn}(k)$ is the assortativity. $\epsilon$ is the efficiency. $Q$ is the weighted modularity of the network.}
   \label{TableCharacteristcs}     
      \begin{tabular}{lcccccccccc}
      \hline 
    Network & $N$ & $|E|$ & $<k>$  & $\omega $  & $\zeta$ & $k_{nn}(k)$ & $\epsilon$ & $Q$\\
    \hline

    Zachary's Karate Club & 33 & 77 & 13.59 & 0.139 & 0.256 & -0.476  & 0.492 & 0.444\\

    Wind Surfers & 43 & 336 & 56.09 & 0.372 & 0.564 & -0.147 & 0.679  & 0.371\\
    
    Madrid Train Bombing & 62 & 243 & 8.81 & 0.121 & 0.561 & 0.029 & 0.448 & 0.435 \\
    
    Les Mis\'erables & 77 & 254 & 21.30 & 0.087 & 0.499 & -0.165 & 0.435 & 0.565 \\

    Wiki Science & 687 & 6,523 & 7.35 & 0.028 & 0.469 & 0.244 & 0.323 & 0.631 \\
    
    Unicode Languages & 868 & 1,255 & 0.697 & 0.003 & 0.00 & -0.171 & 0.255 & 0.772 \\
    
    Scientific Collaboration & 16,726 & 47,594 & 9.23 & 0.0003 & 0.360 & 0.185 & 0.117 & 0.873\\
    \hline
      \end{tabular}
\end{table}

\section{Evaluation Measures}
\label{sec:evaluationmeasures}
We compare the effectiveness of the proposed backbone extraction technique and the ``overlapping hub ego backbone" based on four different evaluation measures classically used. \newline
\textbf{1. Average weighted degree:} The weighted degree of a node is the sum of the weights of all the edges connected to it. Hence, a higher average weighted degree backbone means that important nodes are kept in the graph, reflecting its connectedness. It is defined as follows:

\begin{equation}
  <k> = \frac{1}{N}\sum_{i=1}^N k_i = \sum_{j \in \mathcal{N}(1)} w_{ij}
\end{equation}
where $\mathcal{N}(1)$ is the first-order neighborhood of node $i$. \newline
\textbf{2. Average link weight:} Links in the backbone preserve the information flow of the network. In other words, the higher the value of the links, the better the backbone in maintaining the core information of the graph. It is defined as follows:
\begin{equation}
  <w> = \frac{1}{N}\sum_{i,j \in V} w_{ij}
\end{equation}
\newline
\textbf{3. Average betweenness:} Nodes with higher betweenness can disseminate information quickly. Hence, a backbone with higher average betweenness indicates that the speed of information dissemination is barely altered. It is defined as follows:
\begin{equation}
  <b> = \frac{1}{N}\sum_{i=1}^N b_i = \sum_{i \neq s \neq t} \frac{\sigma_{s,t}^i}{\sigma_{s,t}}
\end{equation}

where $\sigma_{s,t}$ denotes the number of shortest paths between nodes $s$ and $t$ and $\sigma_{s,t}^i$ denotes the number of shortest paths between nodes $s$ and $t$ passing through node $i$. \newline
\textbf{4. Weighted Modularity:} Modularity assesses the quality of the community structure based on the difference between the actual and the expected fraction of edges in the communities. A backbone with higher modularity suggests that the community structure is less altered. It can be computed on unweighted and weighted networks \cite{newman2004analysis}. It is defined as follows: 
\begin{equation}
  Q = \frac{1}{2|E|} \sum_{i,j}\left[ A_{ij} - \frac{w_iw_j}{\sum_{i,j}w_{ij}}\right] \delta(c_i,c_j)
\end{equation}
where $A_{ij}$ is the weighted adjacency matrix of graph $G$ and $\delta(c_i,c_j)$ equals 1 if nodes $i$ and $j$ belong to the same community, otherwise it equals 0.

\section{Experimental Results}
\label{sec:experimentalresults}
The effectiveness of ``modularity vitality backbone" is compared with another community-aware backbone extraction technique recently introduced \cite{ghalmane2020extracting}. The authors propose two backbones in their work, namely ``overlapping nodes ego backbone" and ``overlapping nodes and hubs backbone." In their comparative evaluation, they show that both techniques perform favorably compared to the popular disparity filter. 

Therefore, in this work, we restrict our comparison to the most effective: overlapping nodes ego backbone. Table \ref{TableResults} reports the experimental results for seven real-world networks under study. The backbones quality measures are the average weighted degree ($<k>$), the average link weight ($<w>$), the average betweenness ($<b>$), and the weighted modularity ($Q$). As in their paper, we fix the backbone size to 30\% of the original network.

Let's first discuss the average weighted degree. The higher its value, the better the backbone is in keeping the salient nodes maintaining its connectedness. Table \ref{TableResults} reports that the modularity vitality backbone outperforms overlapping nodes ego backbone in all of the networks under study. Moreover, the difference ranges from very small magnitudes to orders of magnitude higher. For example, in Wiki Science, the average weighted degree in the modularity vitality backbone is eleven times higher than the ``overlapping nodes ego backbone." On the contrary, in Scientific Collaboration, the difference is barely noticeable (0.02).

Let's turn to average link weight. The average link weight characterizes the relevance of the links kept in a backbone. Hence, the higher its value, the better the backbone is in preserving essential links. The results show that the modularity vitality backbone outperforms the overlapping nodes ego backbone in six out of the seven networks. 

Now, we discuss the average betweenness. This measure indicates the amount of information flow that can pass through the nodes of a given backbone. The higher its value, the higher the efficiency of the backbone in information spreading. The modularity vitality backbone outperforms the overlapping nodes ego backbone in only one out of the seven networks. In Scientific Collaboration networks, their values are comparable. It indicates that the information spread within the modularity vitality backbone is not as efficient as overlapping nodes ego backbone. Note, however, that the differences between the two backbones are less pronounced.

Finally, we turn to weighted modularity. The higher the modularity of the backbone, the better the quality of its community structure. As reported in table \ref{TableResults}, the modularity vitality backbone outperforms the overlapping nodes ego backbone on all the networks under study. These results are not surprising. Indeed, the modularity vitality backbone prunes the nodes contributing less to the modularity of the network. Hence, it tends to preserve the modularity as pruning proceeds.

To summarize, the modularity vitality backbone exhibits a higher weighted modularity than the overlapping nodes ego backbone. It preserves essential nodes in the network (i.e., hubs and bridges), thus achieving a higher average weighted degree and average link weight. However, maintaining the community structure comes at a price of a lower average betweenness.
 
\begin{table}[t]
\centering
\caption{The computed values for the average weighted degree ($<k>$), link weight ($<w>$), and betweenness ($<b>$) alongside the weighted modularity ($Q$) of the backbones with 30\% of the initial size of the network. For brevity, MV stands for ``modularity vitality backbone," and OE stands for ``overlapping nodes ego backbone."}
\begin{tabular}{l|ll|ll|ll|ll} 
\hline
\multirow{2}{*}{Network} & \multicolumn{2}{l|}{$<k>$} & \multicolumn{2}{l|}{$<w>$} & \multicolumn{2}{l|}{$<b>$} & \multicolumn{2}{l}{$Q$}  \\ 
\cline{2-9}
                                          & MV & OE                & MV & OE                & MV & OE               & MV & OE                 \\ 
\hline
Zachary's Karate Club                     & \textbf{13.00}  & 8.40        & \textbf{6.05}  & 4.07           & 0.12  & \textbf{0.27}        & \textbf{0.35}  & 0.32                 \\
Wind Surfers                            & \textbf{71.38}   & 35.08        & \textbf{35.69}  & 17.54                & 0.11  & \textbf{0.15}                & \textbf{0.36}  & 0.32                \\
Madrid Train Bombing                  & \textbf{8.53} & 3.90           & \textbf{4.26}   &  1.95               & 0.09    & \textbf{0.14}                 &  \textbf{0.38} &  0.17                  \\
Les Mis\'erables                             &  \textbf{39.48}  & 19.08        & \textbf{19.74}    &  9.54        & 0.08   & \textbf{0.14}               &  \textbf{0.49}  &  0.48                   \\
Wiki Science                            &  \textbf{10.16}  &  0.92                &    \textbf{5.08} &   0.46              &  0.01  & \textbf{0.08}                & \textbf{0.73}   & 0.72               \\
Unicode Languages                       & \textbf{1.46}  &  1.28                &  \textbf{0.73}  &   0.64               &   \textbf{0.03} & 0.02                  &   \textbf{0.79} &  0.78                 \\

Scientific Collaboration                & \textbf{17.22}   &  17.20                 &   4.71 &    \textbf{8.60}              &   0.001 & 0.001               & \textbf{0.81}   & 0.71                \\
\hline
\end{tabular}
\label{TableResults}
\end{table}

\section{Discussion}
\label{sec:discussion}
The constant increase of real-world networks size has prompted researchers to design a smaller yet accurate representation of networks. This problem is tackled either with coarse-graining or filter-based methods. A recent work by Ghalmane et al. \cite{ghalmane2020extracting} has shown interest in exploiting the modular structure of the network to deal with this issue. Building on this finding, we propose a new backbone extractor, ``modularity vitality backbone," that aims to preserve the quality of the community structure. Assigning a modularity vitality score to the nodes, it prunes those with a low contribution to the network's modularity. 

We performed a comparative analysis with the recently introduced ``overlapping nodes ego backbone. These investigations on a set of seven real-world networks from various domains are globally at the advantage of the proposed technique. After pruning 70\% of the network, results show that the modularity vitality backbone maintains higher modularity than overlapping hubs ego backbone. This expected behavior comes with higher performance in terms of average node degree and average link weight. Nonetheless, information efficiency isn't guaranteed. Indeed, one can point out that in five out of seven networks, the modularity vitality backbone suffers from lower information efficiency.

It can be explained by how the modularity vitality backbone proceeds. Indeed, one removes nodes that barely affect modularity. Those nodes may have high betweenness, yet they do contribute much to the network's modular structure. If we consider the overlapping nodes ego backbone, it appears that nodes with high betweenness tend to be overlapping nodes or nodes near the overlaps. Consequently, they are preferred and kept in the backbone.
Nonetheless, they may not contribute to the modularity of the network as other less influential nodes. Another distinction lies in the fact that the modularity vitality backbone doesn't remove edges with low weights. On average, it has a higher number of links compared to the overlapping nodes ego backbone. Thus, it is normal to have lower average betweenness values due to the existence of those edges that play a role in maintaining the network's modularity. In other words, they play a role in showing a clearly defined community structure.

\begin{figure}[!ht]
\centering
\includegraphics[width=5in, height=4.5 in]{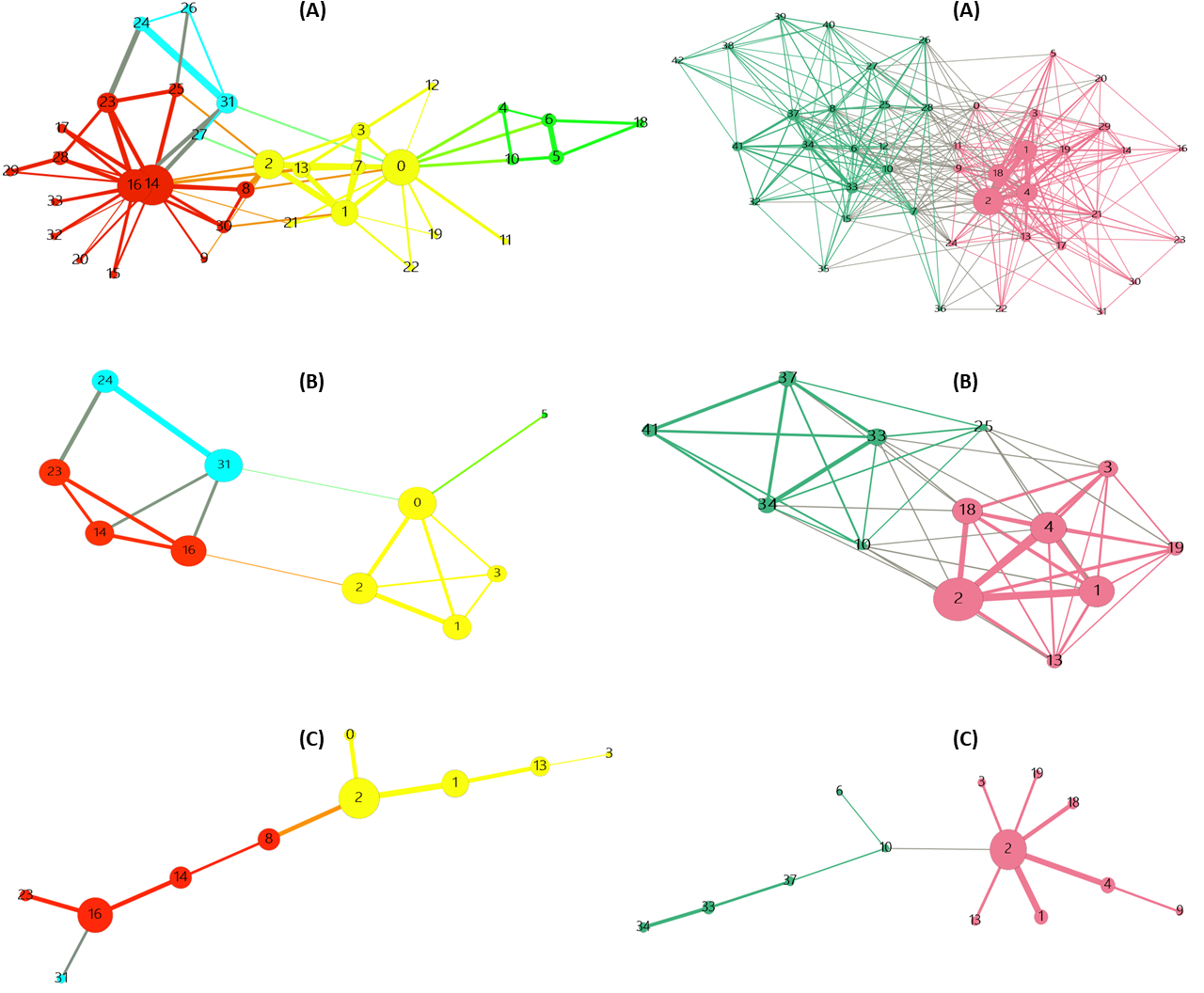}
\caption{The backbone extraction of two networks: Zachary Karate Club on the left and Wind Surfers on the right. (A) represents the original network. (B) represents the modularity vitality backbone. (C) represents the overlapping nodes ego backbone. The different colors of the nodes correspond to the various communities uncovered using the Louvain community detection algorithm. The size of the nodes is proportional to their weighted degree. The size of the edges is proportional to their edge weight.}
\label{Fig1}
\end{figure}

To illustrate these differences, we refer to two small networks, namely Zachary Karate Club and Wind Surfers, given in Figure \ref{Fig1}. Let’s discuss Zachary Karate Club first, represented on the left side of Figure \ref{Fig1}.  One can point out that the modularity vitality backbone represented in (B) has a clique-like topology. In contrast, the overlapping nodes ego backbone reported in (C)  exhibits a star-like structure. It is a good illustration of why the modularity vitality backbone has higher modularity while ``overlapping nodes ego backbone” is characterized by higher information spreading efficiency. Moreover, one can note that the number of communities is the same as the original network using the proposed technique. In contrast, there are no more nodes from the green community in the overlapping nodes ego backbone. 
Diving deeper, the modularity vitality backbone keeps node 24 in the blue community and node 5 in the green community. They are discarded by overlapping nodes ego backbone and replaced by node 13 from the yellow community and node 8 from the red community. If we look at the modularity vitality and betweenness scores of these nodes, we find that nodes 24 and 5 have high modularity but low betweenness. In contrast, nodes 13 and 8 have high betweenness and low modularity vitality scores.  

The Wind Surfers network exhibits similar behavior. Indeed, the modularity vitality backbone shows a clear clique-like structure while a star-like structure emerges in the ``overlapping nodes ego backbone.” Overall, the modularity vitality backbone integrates more peripherical nodes while the overlapping nodes ego backbone tends to retain more nodes at the core of the communities.

Note that the proposed backbone extraction process can also integrate a further step reducing the number of edges. One may consider several strategies to do so. For instance, one may remove the links based on their weights in each community, preserving its connectedness. Another approach is to use the disparity filter to prune these edges. Additionally, one may prune edges in proportion to the size of the edge set in each community of the original network. So doing allows better preservation of the original community structure.
Additionally, preserving the nodes in the backbone according to the absolute value of the modularity vitality scores can be too brutal. One may integrate more information about the community structure, such as the community size, to better deal with the resolution limit issue \cite{fortunato2007resolution}.



\section{Conclusion}
\label{sec:conc}
Analyzing large-scale networks is essential to characterize their underlying topology and dynamics. However, the large size of networks hinders this process. Therefore, it is vital to remove redundant information from the network while keeping nodes and edges that preserve the relevant information. Backbones, whether coarse-grained or filter-based, tackle this problem. 

Aware of the ubiquity of the modular structure of real-world networks, we propose a filter-based technique called ``modularity vitality backbone." The proposed algorithm aims to preserve the network's modularity as nodes are removed. This enables researchers to conduct studies on networks with smaller sizes yet maintain their dense regions, which in turn represent the main building blocks of the network. The proposed method extracts the backbone of real-world weighted networks after quantifying the contribution of the nodes to the overall modularity of the network. Based on these scores, one prunes the nodes that barely contribute to the network's modularity until one reaches the desired size of the backbone.

Experiments show that the modularity vitality backbone compares favorably with its alternative in terms of weighted modularity, average weighted degree, and average link weight. However, it doesn't necessarily keep the nodes contributing to the efficiency of information spreading. Instead, it preserves the nodes and their edges that strategically contribute to the modularity of the network. These results pave the way to developing a filtering backbone extractor dedicated to optimizing several quality measures simultaneously using the vitality framework.

In the short term, we plan to extend this preliminary work in various directions. Since modularity has known drawbacks, we plan to evaluate alternative mesoscopic quality measures. Moreover, we will develop the analysis using multiple mesoscopic and macroscopic evaluation measures. Additionally, we aim to investigate the influence of community detection algorithms on the backbone extraction process.

%
%

\bibliographystyle{unsrt}
\bibliography{bibtech}

\begin{thebibliography}{10}

\bibitem{zeng2011coarse}
An~Zeng and Linyuan L{\"u}.
\newblock Coarse graining for synchronization in directed networks.
\newblock {\em Physical Review E}, 83(5):056123, 2011.

\bibitem{zeng2019new}
Lang Zeng, Zhen Jia, and Yingying Wang.
\newblock A new spectral coarse-graining algorithm based on k-means clustering
  in complex networks.
\newblock {\em Modern Physics Letters B}, 33(01):1850421, 2019.

\bibitem{coscia2017backboning}
Michele Coscia and Frank Neffke.
\newblock Network backboning with noisy data.
\newblock In {\em International Conference on Data Engineering (ICDE)}, 2017.

\bibitem{10.1093/comnet/cnab021}
Tiago Simas, Rion~Brattig Correia, and Luis~M Rocha.
\newblock {The distance backbone of complex networks}.
\newblock {\em Journal of Complex Networks}, 9(6), 10 2021.

\bibitem{boccaletti2006complex}
Stefano Boccaletti, Vito Latora, Yamir Moreno, Martin Chavez, and D-U Hwang.
\newblock Complex networks: Structure and dynamics.
\newblock {\em Physics reports}, 424(4-5):175--308, 2006.

\bibitem{jebabli2018community}
Malek Jebabli, Hocine Cherifi, Chantal Cherifi, and Atef Hamouda.
\newblock Community detection algorithm evaluation with ground-truth data.
\newblock {\em Physica a: Statistical mechanics and its applications},
  492:651--706, 2018.

\bibitem{peel2020detectability}
Leto Peel and Michael~T Schaub.
\newblock Detectability of hierarchical communities in networks.
\newblock {\em arXiv preprint arXiv:2009.07525}, 2020.

\bibitem{atzmueller2021mining}
Martin Atzmueller, Stephan G{\"u}nnemann, and Albrecht Zimmermann.
\newblock Mining communities and their descriptions on attributed graphs: a
  survey.
\newblock {\em Data Mining and Knowledge Discovery}, 35(3):661--687, 2021.

\bibitem{newman2006modularity}
Mark~EJ Newman.
\newblock Modularity and community structure in networks.
\newblock {\em Proceedings of the national academy of sciences},
  103(23):8577--8582, 2006.

\bibitem{ghalmane2020extracting}
Zakariya Ghalmane, Chantal Cherifi, Hocine Cherifi, and Mohammed El~Hassouni.
\newblock Extracting backbones in weighted modular complex networks.
\newblock {\em Scientific Reports}, 10(1):1--18, 2020.

\bibitem{ghalmane2021extracting}
Zakariya Ghalmane, Chantal Cherifi, Hocine Cherifi, and Mohammed El~Hassouni.
\newblock Extracting modular-based backbones in weighted networks.
\newblock {\em Information Sciences}, 576:454--474, 2021.

\bibitem{van2012high}
Martijn~P Van Den~Heuvel, Ren{\'e}~S Kahn, Joaqu{\'\i}n Go{\~n}i, and Olaf
  Sporns.
\newblock High-cost, high-capacity backbone for global brain communication.
\newblock {\em Proceedings of the National Academy of Sciences},
  109(28):11372--11377, 2012.

\bibitem{cao2019motif}
Jie Cao, Cuiling Ding, and Benyun Shi.
\newblock Motif-based functional backbone extraction of complex networks.
\newblock {\em Physica A: Statistical Mechanics and its Applications},
  526:121123, 2019.

\bibitem{zweig2011systematic}
Katharina~Anna Zweig and Michael Kaufmann.
\newblock A systematic approach to the one-mode projection of bipartite graphs.
\newblock {\em Social Network Analysis and Mining}, 1(3):187--218, 2011.

\bibitem{neal2013identifying}
Zachary Neal.
\newblock Identifying statistically significant edges in one-mode projections.
\newblock {\em Social Network Analysis and Mining}, 3(4):915--924, 2013.

\bibitem{neal2014backbone}
Zachary Neal.
\newblock The backbone of bipartite projections: Inferring relationships from
  co-authorship, co-sponsorship, co-attendance and other co-behaviors.
\newblock {\em Social Networks}, 39:84--97, 2014.

\bibitem{gfeller2007spectral}
David Gfeller and Paolo De~Los~Rios.
\newblock Spectral coarse graining of complex networks.
\newblock {\em Physical review letters}, 99(3):038701, 2007.

\bibitem{chen2016effectively}
Mei Chen, Longjie Li, Bo~Wang, Jianjun Cheng, Lina Pan, and Xiaoyun Chen.
\newblock Effectively clustering by finding density backbone based-on knn.
\newblock {\em Pattern Recognition}, 60:486--498, 2016.

\bibitem{serrano2009extracting}
M~{\'A}ngeles Serrano, Mari{\'a}n Bogun{\'a}, and Alessandro Vespignani.
\newblock Extracting the multiscale backbone of complex weighted networks.
\newblock {\em Proceedings of the national academy of sciences},
  106(16):6483--6488, 2009.

\bibitem{goh2006skeleton}
K-I Goh, Giovanni Salvi, Byungnam Kahng, and Doochul Kim.
\newblock Skeleton and fractal scaling in complex networks.
\newblock {\em Physical review letters}, 96(1):018701, 2006.

\bibitem{zhang2018extracting}
Ronda~J Zhang, H~Eugene Stanley, and Y~Ye Fred.
\newblock Extracting h-backbone as a core structure in weighted networks.
\newblock {\em Scientific reports}, 8(1):1--7, 2018.

\bibitem{DAI2018271}
Liang Dai, Ben Derudder, and Xingjian Liu.
\newblock Transport network backbone extraction: A comparison of techniques.
\newblock {\em Journal of Transport Geography}, 69:271--281, 2018.

\bibitem{RePEc}
Kanokwan Malang, Shuliang Wang, Yuanyuan Lv, and Aniwat Phaphuangwittayakul.
\newblock {Skeleton Network Extraction and Analysis on Bicycle Sharing
  Networks}.
\newblock {\em International Journal of Data Warehousing and Mining (IJDWM)},
  16(3):146--167, July 2020.

\bibitem{koschutzki2005centrality}
Dirk Kosch{\"u}tzki, Katharina~Anna Lehmann, Leon Peeters, Stefan Richter,
  Dagmar Tenfelde-Podehl, and Oliver Zlotowski.
\newblock Centrality indices.
\newblock In {\em Network analysis}, pages 16--61. Springer, Berlin,
  Heidelberg, 2005.

\bibitem{yang2015defining}
Jaewon Yang and Jure Leskovec.
\newblock Defining and evaluating network communities based on ground-truth.
\newblock {\em Knowledge and Information Systems}, 42(1):181--213, 2015.

\bibitem{leskovec2010empirical}
Jure Leskovec, Kevin~J Lang, and Michael Mahoney.
\newblock Empirical comparison of algorithms for network community detection.
\newblock In {\em Proceedings of the 19th international conference on World
  wide web}, pages 631--640, 2010.

\bibitem{rajeh2021characterizing}
Stephany Rajeh, Marinette Savonnet, Eric Leclercq, and Hocine Cherifi.
\newblock Characterizing the interactions between classical and community-aware
  centrality measures in complex networks.
\newblock {\em Scientific Reports}, 11(1):1--15, 2021.

\bibitem{clauset2004finding}
Aaron Clauset, Mark~EJ Newman, and Cristopher Moore.
\newblock Finding community structure in very large networks.
\newblock {\em Physical review E}, 70(6):066111, 2004.

\bibitem{brandes2007modularity}
Ulrik Brandes, Daniel Delling, Marco Gaertler, Robert Gorke, Martin Hoefer,
  Zoran Nikoloski, and Dorothea Wagner.
\newblock On modularity clustering.
\newblock {\em IEEE transactions on knowledge and data engineering},
  20(2):172--188, 2007.

\bibitem{blondel2008fast}
Vincent~D Blondel, Jean-Loup Guillaume, Renaud Lambiotte, and Etienne Lefebvre.
\newblock Fast unfolding of communities in large networks.
\newblock {\em Journal of statistical mechanics: theory and experiment},
  2008(10):P10008, 2008.

\bibitem{chen2015fuzzy}
Mingming Chen and Boleslaw~K Szymanski.
\newblock Fuzzy overlapping community quality metrics.
\newblock {\em Social Network Analysis and Mining}, 5(1):1--14, 2015.

\bibitem{magelinski2021measuring}
Thomas Magelinski, Mihovil Bartulovic, and Kathleen~M Carley.
\newblock Measuring node contribution to community structure with modularity
  vitality.
\newblock {\em IEEE Transactions on Network Science and Engineering},
  8(1):707--723, 2021.

\bibitem{newman2004analysis}
Mark~EJ Newman.
\newblock Analysis of weighted networks.
\newblock {\em Physical review E}, 70(5):056131, 2004.

\bibitem{fortunato2007resolution}
Santo Fortunato and Marc Barthelemy.
\newblock Resolution limit in community detection.
\newblock {\em Proceedings of the national academy of sciences}, 104(1):36--41,
  2007.

\end{thebibliography}

\end{document}